\documentclass[prl,twocolumn,showpacs,superscriptaddress]{revtex4-2}
\usepackage{graphicx,epstopdf}
\usepackage{amsmath}
\usepackage{amssymb}
\usepackage{color}
\usepackage{float}
\usepackage{epsfig}
\usepackage{epstopdf}
\usepackage{framed}
\usepackage{subfigure}
\usepackage{enumerate}

\usepackage[colorlinks=true, citecolor=blue, urlcolor=blue ]{hyperref}

\newcommand{\wt}{\omega_\mathrm{T}}

\newcommand{\old}[1]{{\color[rgb]{0,0.8,0}{..}}}
\newcommand{\floold}[1]{{\color[rgb]{0,0,0.8}{..}}}

\newcommand{\Imperial}{
Physics Department, Blackett Laboratory, Imperial College London, Prince Consort Road, SW7 2AZ, United Kingdom}
\newcommand{\IITB}{Department of Physics, Indian Institute of Technology Bombay, Powai, Mumbai 400076, India}

\begin{document}

\title{Quest for Vortices in Photon Condensates}

\author{Himadri S. Dhar}\email{himadri.dhar@iitb.ac.in}\affiliation{\Imperial}\affiliation{\IITB}
\author{Zai Zuo} \affiliation{\Imperial}
\author{Jo\~ao D. Rodrigues}\affiliation{\Imperial}
\author{Robert A. Nyman}\affiliation{\Imperial}
\author{Florian Mintert}\affiliation{\Imperial}

\begin{abstract}
We 
predict that a photon condensate inside a dye-filled microcavity forms long-lived spatial structures that resemble vortices when incoherently excited by a focused pump orbiting around the cavity axis. The finely structured density of the condensates have a discrete rotational symmetry that is controlled by the orbital frequency of the pump spot and is phase-coherent over its full spatial extent despite the absence of any effective photon-photon interactions. 
\end{abstract}

\maketitle


Vortices are a ubiquitous phenomenon occurring in a broad range of many-body systems, wherein a robust topological defect prevents a 
phase singularity at the middle of a circulating pattern. 
They appear in fluid turbulence~\cite{Falkovich2018}, magnetic structures in thin films~\cite{Wachowiak2002}, 
superconductors~\cite{Blatter1994}, and atomic condensates~\cite{Ketterle2002}. Formation of vortices in a Bose-Einstein condensate (BEC) is ultimately related to superfluidity~\cite{Leggett1999,Carusotto2013}, which is a hallmark of phase coherence in quantum systems that arises from interactions between the particles.
While quantum coherence is an integral part of conservative quantum systems, the existence of superfluidity and vortices in driven-dissipative systems such as polariton condensates~\cite{Lagoudakis2008, Amo2009} is remarkable. It demonstrates that long-range coherence persists beyond typical loss timescales of the system due to the macroscopic nature of condensation~\cite{Kasprzak2006}. Vortices in BECs may be generated either by flowing the fluid past a static obstacle~\cite{Amo2009},  %
colliding two condensates~\cite{Rodrigues2020} or by stirring the potential landscape~\cite{Abo2001}.

%

In recent years, a key development in condensate physics has been the creation of driven-dissipative Bose-Einstein condensates (BECs) of photons in microcavities filled with a flourescent dye \cite{Klaers2010b,Marelic2015,Greveling2018}. In contrast to conservative and polaritonic BECs, no superfluidity has so far been observed in photon condensates, because there is no significant photon-photon interaction \cite{Alaeian2017,Nyman2014}.
Similar to thermalization, which is achieved through interaction with the dye molecules, also coherence in the condensed light builds up only on account of the emission from the dye molecules~\cite{Snoke2013}. While this mechanism is sufficient to result in the establishment of long-range phase coherence in a {photon BEC}~\cite{Schmitt2016,Marelic2016,Damm2017}, clear signatures of superfluidity such as the formation of vortices have remained elusive.

Our main goal is to {theoretically} explore to what extent a transient photon BEC can exhibit macroscopic coherence resembling that of a vortex. 
Rather than rotationally deforming the trapping potential, the dynamics in the condensate is 
driven by an orbiting pump spot. 
Stirring is thus achieved by an incoherent, and not a coherent mechanism.
The resulting dynamical features share striking similarities with vortices in {conservative BECs, which arise from superfluidity, but with some fundamental differences}. For instance, 
the condensate forms a rigid spatial structure that rotates with the orbital frequency of the pump spot and has a high-degree of phase coherence.
However, in contrast to {vortices in conservative condensates that have a density with continuous {rotational symmetry}},
the photon BEC adopts a finely structured density with only a discrete rotational symmetry
and an order that is determined by the {orbital} frequency of the pump spot.

\begin{figure}[t]
\epsfig{figure = 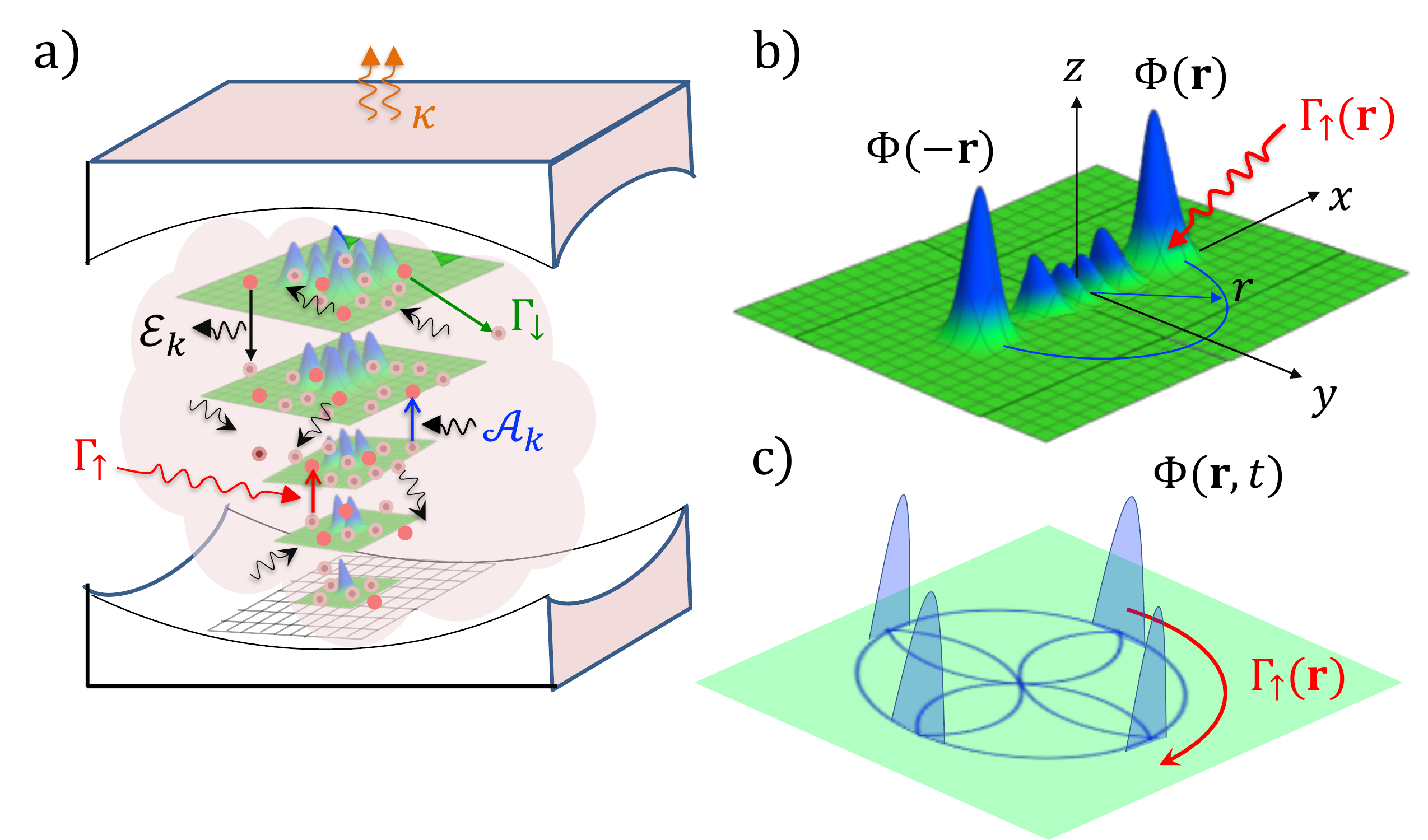, width=0.48\textwidth,angle=-0}
\caption{(Color online.) a) An illustration of a dye-filled microcavity, showing the first few nondegenerate cavity mode intensities, $|\psi_k(\mathbf{r})|^2$, on the 2D transverse plane. The absorption and emission rates for cavity mode $k$ are given by $\mathcal{A}_k$ and $\mathcal{E}_k$, and $\kappa$ is the rate of photon loss.  $\Gamma_\uparrow$ is the pump rate and $\Gamma_\downarrow$ is the rate at which molecular excitation is lost to non-cavity modes.
b) The spatially resolved photon wavepacket, $\Phi(\mathbf{r})$, oscillating on the transverse $xy$ plane, across the vertical cavity axis $z$, for an external pump focused at $\mathbf{r}$. c) A qualitative picture of peaks forming during the time-evolution of $\Phi(\mathbf{r},t)$, due to periodic coming together of the oscillating wavepacket in the harmonic trap and the orbiting pump.
}
 \label{fig0}
\end{figure}

{In a typical setup,} the photon gas inside the microcavity is restricted to a single longitudinal mode, 
denoted by the cavity cutoff frequency $\omega_0$, while the mirror curvature imposes a harmonic potential on the two-dimensional (2D) transverse plane. The energy and wavefunction of the transverse cavity mode $k$ is given by $\omega_k$ and $\psi_k(\mathbf{r})$. 
Now, an incoherent external pump with rate $\Gamma_\uparrow$ and focused on a fixed region or spot on the transverse plane, 
produces a nonequilibrium distribution of excited dye molecules. 
The subsequent 
behavior of the emitted photons is dependent on the system properties, such as rate of absorption $\mathcal{A}_k$ and emission $\mathcal{E}_k$, and photon loss rate $\kappa$, as illustrated in Fig.~\ref{fig0}(a). Importantly, the thermalization of photons is directly related to the total number of absorptions per unit photon loss, 
while the condensation transition is controlled by the pump rate.
For an off-center pump, a transient photon wavepacket is formed close to the pump spot (see Fig.~\ref{fig0}(b)). Under conditions for good thermalization, the photon wavepacket collapses to the center to form a near-equilibrium Bose-Einstein condensate, whereas for poor thermalization, the stimulated light oscillates inside the harmonic trap imposed by the cavity mirrors. The formation and kinetics of a nonequilibrium, mode-locked photon wavepacket in an effective one-dimensional space has been experimentally demonstrated using a pulsed pump \cite{Schmitt2015}. {Such transient behavior of light inside the cavity can be closely simulated using a microscopic, nonequilibrium model of photon condensation \cite{Kirton2013,Keeling2016}, which provides us the necessary theoretical tools for the study.}

Photon emission with more enriched spatial features 
{begin to appear} in the transient dynamics when the focused pump spot is no longer static but
orbiting in the transverse plane,
as shown in Fig.~\ref{fig0}(c).
{The dye molecules} located within an annulus of radius $r$ and width $w$ determined by orbit and width of the pump spot are initially excited.
The optical modes that overlap with the annular region compete for these excitations and interfere to produce an initial displaced packet of light through stimulated emission (a non-equilibrium condensate).
The evolution of the emitted photons is also dependent on the thermalization condition.
Specifically, the low thermalization regime (absorption is slow compared to other processes), where nonequilibrium effects dominate, is 
{favorable for the formation of spatial structures that exhibit phase coherence.}

\begin{figure}[t]
\centering
\epsfig{figure = 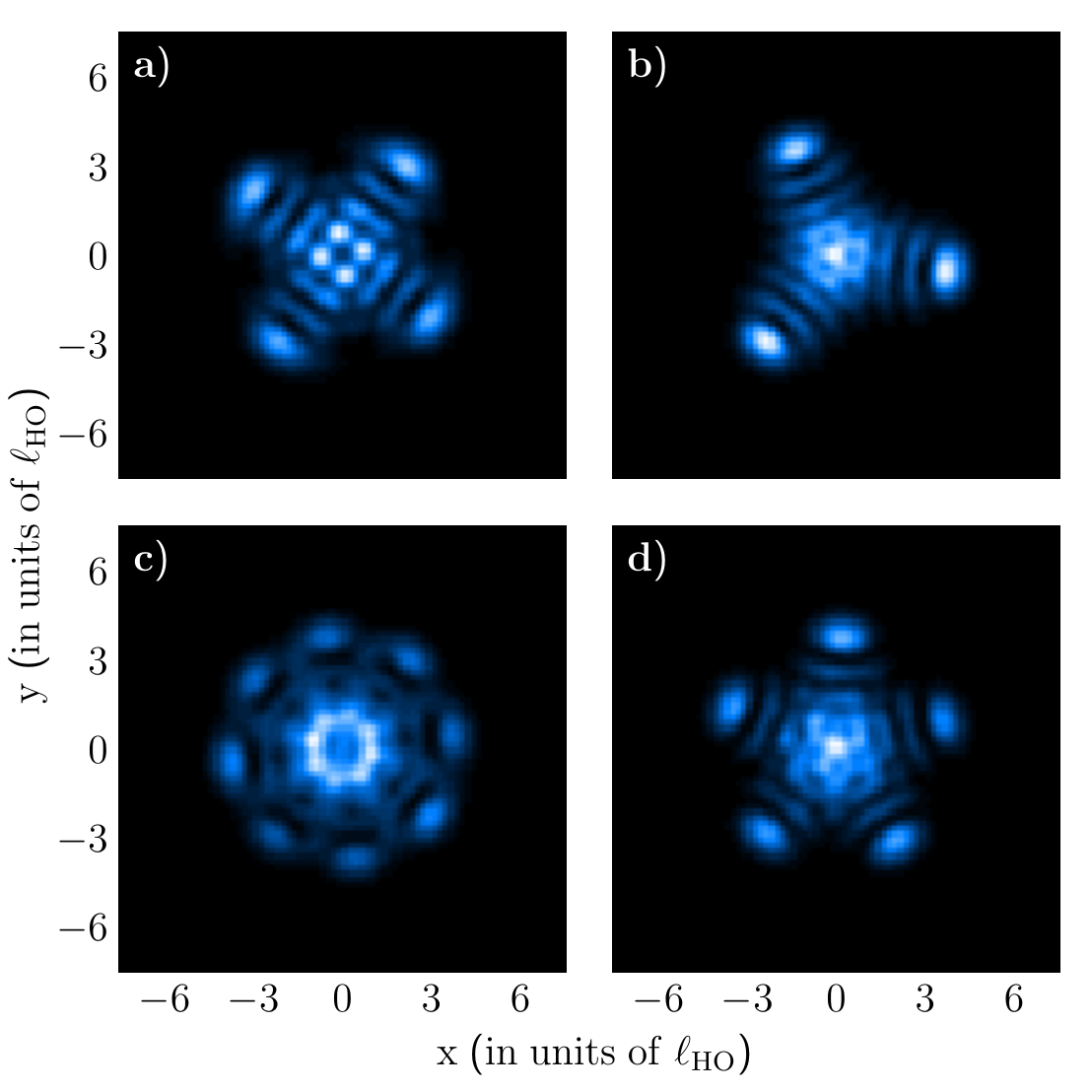, width=0.48\textwidth,angle=-0}
\caption{(Color online.) Photon density $I(\mathbf{r})$ in the transverse plane of the cavity. The figure exhibits a snapshot of the photon emission, when the dye-filled cavity is driven by an incoherent pump, focused at distance $r$ away from the cavity center and circulating around this center with frequency, $\nu =  \wt/z$, where a) $z = 2$, b) $z = 3$, c) $z = 4$, and d) $z = 5$. All the axes are in units of harmonic oscillator length, $\ell_\mathrm{HO}$.
}
 \label{fig1}
\end{figure}

{Shortly after the start of illumination,
the photons
begin to form a ring-shaped structure, which underlines the stimulated emission along the annular region traversed by the pump.
Subsequently, this structure deforms,
and more complex spatial structures in the photon density begin to emerge.
If the orbital frequency $\nu$ of the pump spot is a fraction of the frequency $\wt$ of the harmonic potential that traps the photon gas, i.e., $\nu = \wt/z$, where $z$ is a positive integer,
the photon density evolves towards a polygon pattern with a discrete rotational symmetry. Depending on whether $z$ is odd or even, the polygon has $z$ or $2z$ vertices, respectively and the structure rotates with the orbital frequency $\nu$ of the pump spot.}
As the system evolves, the edges of the polygon disappear and a {rigid} structure with discrete peaks at the vertices is formed, with interference between the peaks. 
Such {spatial structures}, for both odd and even values of $z$,  are shown in Fig.~\ref{fig1} in terms of the photon density
$I(\mathbf{r},t) = \sum_{k,k'} \Psi_{k,k'}(\mathbf{r}) ~n_{k,k'}(t)$ 
on the transverse plane of the cavity.
Here, $\Psi_{k,k'}(\mathbf{r})$ = $\psi^*_{k}(\mathbf{r}) \psi_{k'}(\mathbf{r})$ and ${n}_{k,k'}$ = $\langle \hat{a}_k^\dag\hat{a}_{k'}\rangle$, are the mode overlap function and the photon correlation, respectively. 
%
{Once the spatial structure is formed it is robust and long-lived and becomes a characteristic feature of the condensate. For instance, for $\nu = \wt/2$, the intensity of the peaks and symmetric structure are visible even after more than 100 orbital periods have passed since the formation of the condensate.}
Further snapshots of the photon density at various moments during the evolution are given in the Supplemental Material (SM).

The formation of the polygon structure and the discrete peaks can be qualitatively explained in terms of the motion of the photon wavepacket arising from oscillations inside the harmonic trap and the motion of the focused pump spot. 
{Once a wavepacket is created as result of the pump, it will start oscillating in the trapping potential with a maximal displacement given by the radius $r$ of the 
pump orbit. Due to dissipation, any such wavepacket can persist only if it encounters the pump spot regularly,
and the condition for this to happen is understood more easily by following the system dynamics in the frame co-rotating with the pump.
In this frame, a wavepacket follows curved trajectories as depicted in Fig.~\ref{fig0}(c),
and regular encounters with the pump spot 
{are} possible only if this curved trajectory forms a periodic orbit.
The angle of rotation of the trajectory between instances of maximal displacement is given by $\pi(1-1/z)$,
comprised of a contribution of $\pi$ of the wavepacket in the lab-frame and a contribution of $\pi/z$ of the pump.
Thus, the angular rotation of the wavepacket after $z$ instances of maximal displacement is given by $\pi(z-1)$,
which, in the case of odd $z$, is an integer multiple of $2\pi$.
In this case, the rosetta-shaped orbit of the wavepacket is closed (as sketched in Fig.\ref{fig0}(c))), and it contains $z$ points of maximal displacement, resulting in a $z$-fold symmetry.
In the case of even $z$, on the other hand, it takes $2z$ instances of maximal displacement  before the orbit closes and its symmetry is $2z$-fold.}
{As such the complex structures result from the interplay between the nonequilibrium and driving mechanisms acting upon the photon gas.}

{The transient photon density is numerically estimated 
from a set of nonlinear rate equations that not only describe the dynamics of the photons and the molecules (see SM for details) but also provide a more quantitative approach to understand the spatial structures.}
The equation of motion for the photon correlation matrix $\textbf{n}$ with elements $n_{k,k'} = \langle \hat{a}_k^\dag\hat{a}_{k'}\rangle$ is given by
%
\begin{equation}
\label{eq:ndot}
\dot{\textbf{n}} =\left(i\mathbf{\Omega}-\frac{\kappa}{2}\right) \textbf{n} + \rho_0 \{\textbf{f}~\textbf{E}(\textbf{n}+\mathbb{I}) + (\textbf{f}-\mathbb{I})\textbf{A}^\dag\textbf{n}\}+ \mathrm{h.c.}\ ,
\end{equation}
where $\textbf{f}$ is the molecular excitation matrix with elements ${f}_{k,k'}$ = $\sum_{i}\Psi_{k,k'}(\mathbf{r_i}) \langle\sigma_i^+\sigma_i^-\rangle$ and $\rho_0$ is the molecular density.
The absorption and emission matrices are given by ${E}_{k,k'} = \mathcal{E}_k\delta_{k,k'}$ and ${A}_{k,k'} = \mathcal{A}_k\delta_{k,k'}$, respectively. Moreover, 
${\Omega}_{k,k'} = \omega_k \delta_{k,k'}$, where $\omega_k$ is the energy of the cavity mode $k$, and $\mathbb{I}$ is the identity matrix.
The equation of motion for the molecular excitation vector $\mathbf{m}$, with elements $m_j = \sum_{i}\delta(\mathbf{r_j}-\mathbf{r_i})\langle\sigma_i^+\sigma_i^-\rangle$, 
is given by
\begin{eqnarray}
\dot{\mathbf{m}} &=& - \{\Gamma_\downarrow + 2 \mathbf{E}_\mathrm{eff}\}\mathbf{m} 
+\{\Gamma_\uparrow(\mathbf{r}) + 2 \mathbf{A}_\mathrm{eff}\} (\mathbf{1} - \mathbf{m}), \label{eq:fdot}
\end{eqnarray}
where, $\mathbf{E}_\mathrm{eff} = \mathrm{Tr}[\psi(\mathbf{r})\mathbf{E}(\mathbf{n} + \mathbb{I})]$ and
$\mathbf{A}_\mathrm{eff} = \mathrm{Tr}[\psi(\mathbf{r})\mathbf{n}\mathbf{A}]$. 


The motion of the photon wavepacket and the 
formation {of the condensate} can be explained in terms of 
the nonequilibrium dynamics of the system. 
A simplified picture can be constructed by rewriting Eq.~(\ref{eq:ndot}), as $\dot{\textbf{n}} =  \mathbf{\mathcal{F}^{(0)}} + i \mathbf{\mathcal{F}^{(1)}}$, where 
$\mathcal{F}^{(0)}$ is the non-oscillatory component  and $\mathcal{F}^{(1)}$ oscillates with $\Omega_{k,k'}$.
In the absence of $\mathcal{F}^{(1)}$, the rate equation for the photon correlation is $\dot{n}_{k,k'} =  \mathcal{F}_{k,k'}^{(0)}$. For a constant pump at $\mathbf{r}$ and after a long time $\tau$, the photons condense to form a localized wavepacket $\Phi(\mathbf{r},\tau)$, which is the steady state, i.e., $\bar{n}_{k,k'}(\tau)$ = Tr$[\hat{a}^\dag_k \hat{a}_{k'}\Phi(\mathbf{r},\tau)]$ is the steady solution for $\dot{\bar{n}}_{k,k'} =  \mathcal{F}_{k,k'}^{(0)} = 0$.
To approximate the oscillatory dynamics, the term $\mathcal{F}^{(1)}$ is introduced at $t>\tau$ to obtain a new equation of motion
$\dot{n}_{k,k'}  =  i(\omega_k-\omega_{k'})~ \bar{n}_{k,k'}$, with solution $\bar{n}_{k,k'}(t)$ = $\bar{n}_{k,k'}(\tau)\exp[i\Delta\omega_{k,k'}t]$. Here, $\Delta\omega_{k,k'}$ = $\omega_k-\omega_{k'}$,
is the gap between neighboring energy modes of the cavity. 
%
The photon density is then given by
$
I(\mathbf{r}) = \sum_{k,k'} \Psi_{k,k'} \bar{n}_{k,k'}(\tau)\exp[i\Delta\omega_{k,k'}t].
$

Now, for 2D harmonic oscillators, the frequency of mode $k$ is given by $\omega_k = \omega_0 + (q_x+q_y)\wt$, corresponding to the quantum numbers $\{q_x,q_y\}$ and the cutoff frequency $\omega_0$.
Since the pump spot is focused at $r$,
the condensed light is mostly found in the annular domain of radius $r$. 
The highly populated modes thus have a maximum amplitude around this domain.
These modes are characterized by the condition $q_x + q_y$ = $q, q\pm1$, where $q$ is the integer closest  to ${r}^2/(2 \ell_\textrm{HO}^2)$,where $\ell_\textrm{HO}$ is the harmonic oscillator length (see SM for details).
Since degenerate modes do not contribute to the oscillation, as $\Delta\omega_{k,k'} = 0$,
the oscillation comes from nondegenerate modes with $q$ and $q\pm1$, for which
$\Psi_{k,k'}(\mathbf{r}) = \psi^*_{k}(\mathbf{r}) \psi_{k'}(\mathbf{r})$ is an odd function.
At $t = n\pi/\wt$, the density is given by
$I(\mathbf{r}) = \sum_{k,k'} \Psi_{k,k\pm1}(\mathbf{r}) \bar{n}_{k,k\pm1}$,  for even $n$, and by $I(-\mathbf{r}) = -\sum_{k,k'} \Psi_{k,k\pm1}(\mathbf{r}) \bar{n}_{k,k\pm1}$, for odd $n$.
Therefore, the wavepacket  $\Phi(\mathbf{r},\tau)$ oscillates  between the positions $\mathbf{r}$ and $-\mathbf{r}$ in the cavity plane, with frequency $\wt$.
The transient spatial 
{structure} then arises due to the interference between the wave functions of the dominant modes with $q$ and $q\pm1$.

The discrete rotational symmetry with the odd-even dichotomy, as discussed earlier, results from $\Phi(\mathbf{r},\tau)$ being influenced by the two competing frequencies, viz. the harmonic trap frequency $\wt$ and the pump orbital frequency $\nu$.
The constructive interference or beats occur with frequency $\frac{1}{2}(\wt-\nu)$ and yields the factor of $\frac{1}{2}(1-1/z)$ in the angular rotation, which ultimately gives rise to the $z$- or $2z$-fold angular symmetry, as discussed earlier.
The {discrete} symmetry is robust to deviations of $z$ from integer values but, for ostensibly non-commensurate values of the frequencies, the light is spread along the annular region surrounding the pump orbit with no clear spatial 
{structure} \cite{comment1}.

\begin{figure*}[t]
\includegraphics[width=6.6in]{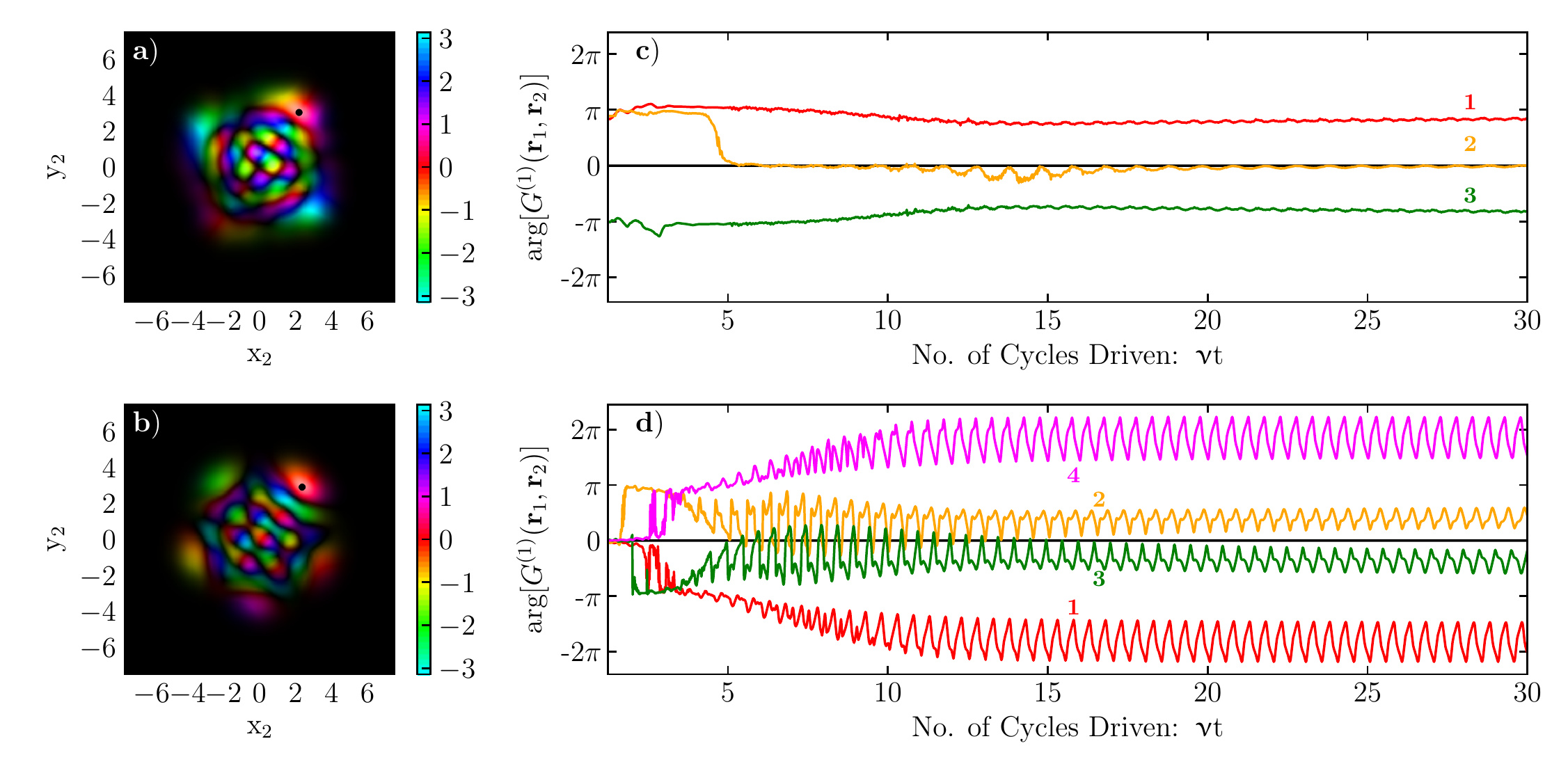}
\caption{First-order correlation function and phase relation in the 
{condensed light.} Subfigures~(a)-(b) show  $G^{(1)}(\mathbf{r}_1,\mathbf{r}_2)$ between a reference point $\mathbf{r}_1$ (shown by a dark dot) and any other spatial point  $\mathbf{r}_2$, corresponding to orbital pump frequency  $\nu = \wt/2$ (inset (a)) and  $\nu = \wt/5$ (inset (b)), in the co-rotating frame. 
The color intensity (dark to bright) indicates the modulus of $G^{(1)}$, while the phase relation given by the argument, is described by the color variation (color bar).
Subfigures (c)-(d) show the unwrapped phase difference between the reference peak at $\mathbf{r}_1$ and the neighboring peaks numbered clockwise as 1-3 (in inset (c)) for 
{structures} with four-fold symmetry, and numbered 1-4 (in inset (d)) for those with five-fold symmetry.
An offset of $-2\pi$ and $2\pi$ is added to peaks 1 and 4, respectively, for ease of viewing.
All the axes in (a)-(b) are in units of harmonic oscillator length, $\ell_\mathrm{HO}$, whereas for (c)-(d) the $x$-axis is the scaled time, $\nu t$, and the $y$-axis is in radians.
}
\label{fig2}
\end{figure*}

The 
{transient structures of the photon density}
depicted in Fig.~\ref{fig1} are consistent with phase coherence in the system, and the explanations given so far are based on the assumption of at least partially coherent dynamics.
{However, unambiguous evidence of phase coherence, can not be obtained from density profiles alone.}
{A proper verification of phase coherence needs to} be obtained in terms of
the first order correlation between photons at different points in the cavity plane. By introducing the field operators $\hat{\xi}(\mathbf{r}) = \sum_k \psi_k(\mathbf{r})\hat{a}_k$ and its conjugate $\hat{\xi}^\dag(\mathbf{r}) = \sum_k \psi_k^*(\mathbf{r})\hat{a}_k^\dag$, the first-order correlation function is defined as, $G^{(1)}(\mathbf{r}_1,\mathbf{r}_2,t)$ = $\langle \hat{\xi}^\dag(\mathbf{r}_1) \hat{\xi}(\mathbf{r}_2)\rangle$ = $\sum_{k,k'} \psi^*_{k}(\mathbf{r}_1) \psi_{k'}(\mathbf{r}_2) ~n_{k,k'}(t)$. 
Figures~\ref{fig2}(a)-(b), shows the correlation function $G^{(1)}(\mathbf{r}_1,\mathbf{r}_2)$ in the co-rotating frame, between a fixed reference point $\mathbf{r}_1$  (shown by a dark dot) and any other point $\mathbf{r}_2$ in the transverse plane. The figures correspond to the orbital pump frequencies $\nu=\wt/z$ for  $z=2$ in inset a) and $z=5$ in inset b).
The color intensity (dark to bright) denotes the modulus of $G^{(1)}(\mathbf{r}_1,\mathbf{r}_2)$, which exhibits either a four-fold or five-fold rotational symmetry, consistent with the photon density of the condensed light for odd and even $z$. The color map highlights the argument of $G^{(1)}(\mathbf{r}_1,\mathbf{r}_2)$, which gives us the phase resulting from the interference between the modes.
{Despite the detailed structure of the photon density with several minima (in black) one can see a clear closed phase evolution for radii of about $3l_{\mbox{\tiny HO}}$ winding around the center. Similar vortex-like phase windings are also observed on a smaller scale.}

The phase has a fixed relation in the co-rotating frame, as seen in Figs.~\ref{fig2}(c)-(d), where 
the phase difference between the spatial points corresponding to the different photon density peaks is depicted as a function of time. 
For $z$ = 2, there are four peaks (numbered 0-3, clockwise) and in the ideal case, it is expected that the phase difference between a reference peak 0 at $\mathbf{r}_1$, and its neighboring peaks 1 and 3 is $\pm\pi$, while it is in phase ($2\pi$) with the diametrically opposite peak 2. This is because for $z$ = 2, the beat phenomenon results in all the peaks being formed in a single pump cycle. Figure~\ref{fig2}(c) shows that the phase correlations converge to these values with time. The situation is more complicated for $z$ = 5, where the five peaks (numbered 0-4, clockwise) are formed over two pump cycles. Hence, taking a reference peak 0 at $\mathbf{r}_1$, the phase difference with the neighboring peak 1 and 4 changes between $\mp\pi/5$ and $\pm3\pi/5$, respectively, and for the distant peaks 2 and 3 between $\pm\pi/5$ and $\pm3\pi/5$, respectively, as seen in Fig.~\ref{fig2}(d). 


Experimental realization of the spatial structures identified here requires 
{an orbiting} pump spot in an otherwise standard room-temperature photon BEC experiment~\cite{Klaers2010}. 
Since this can be implemented by interfering two Laguerre-Gaussian beams of different order and frequency,
experimental observation of these structures would provide very compelling evidence of well-defined coherence properties that can be established despite the absence of any effective particle interaction.
The present results thus not only indicate a pathway towards the observation of vortex-like phenomena in photon BECs, 
but they also support the expectation that features like superfluidity that are common in 
atomic and quasiparticle BECs can also be extended to photons if suitable driving mechanisms are identified.

While the present analysis applies to a regime of large photon numbers, the interference mechanisms resulting in the formation of coherent spatial 
structures
exist for light of any level of intensity, right down to the few-photon regime. 
Since the photon number in a BEC can be controlled in terms of mirror curvature~\cite{Walker2018,Rodrigues2021},
the ability to create well-designed states of light through temporally tuned driving is thus applicable to a broad range of regimes.
Given the notorious difficulty to prepare non-classical few-photon states of light and their potential for quantum-technological applications,
the present work thus may inspire an unconventional handle on the control of quantum states of light.

\begin{acknowledgments}
We acknowledge financial support from the European Commission via the PhoQuS project (H2020-FETFLAG-2018-03) number 820392 and the EPSRC (UK) through the grants EP/S000755/1.
\end{acknowledgments}

\section{Supplemental Material}

In this section, we present some additional information, data and figures that augment the results presented in the main text. We begin with a detailed discussion on the derivation of the equations of motion used to study the dynamics of the photon correlation and the molecular excitation. 
We then highlight the formation of the different spatial structures of the photon density as the system gradually evolves in time. And finally, 
we look at the occupation or population of the different cavity modes that create the interference pattern in the condensed light.


\subsection{I. Mode-coherent nonequilibrium model}
\label{sec0}

To study the spatial characteristics of the photon gas confined inside a dye-filled microcavity, we consider a nonequilibrium model that takes into account  both the population and coherence of the cavity modes~\cite{Keeling2016}.  The master equation describing such a system is given by 
\begin{eqnarray}
{d\rho}/{dt} &=& -i[H_0,\rho] - \mathcal{L}^{(0)}[\rho] - \mathcal{L}^{(1)}[\rho], ~\textrm{where},\\\nonumber\\
\mathcal{L}^{(0)}[\rho] &=& \sum_{k,i} \left\{\kappa {L}(\hat{a}_k) + \Gamma_\uparrow(\mathbf{r_i}) {L}(\hat{\sigma}^+_i) + \Gamma_\downarrow {L}(\hat{\sigma}^-_i) \right\}\rho, \nonumber \\
\mathcal{L}^{(1)}[\rho] &=& \sum_{k,k',i} \Psi_{k,k'}(\mathbf{r_i})  \left\{ \mathcal{A}_{k'} [\hat{a}_{k'}\hat{\sigma}^+_i\rho,\hat{a}^\dag_k\hat{\sigma}^-_i] \right.\nonumber\\
&+&  \left.\mathcal{E}_k [\hat{a}^\dag_k\hat{\sigma}^-_i\rho,\hat{a}_{k'}\hat{\sigma}^+_i]\right\}.
\end{eqnarray}
Here, $H_0 = -\delta_k \hat{a}^\dag_k\hat{a}_k$ is the system Hamiltonian for weak light-matter interaction~\cite{Keeling2016}, where $\delta_k = \omega_\mathrm{zpl}-\omega_k$ is the detuning of the cavity mode frequency $\omega_k$ from the zero-phonon line $\omega_\mathrm{zpl}$. The cavity-cutoff frequency is given by $\omega_0$.
Moreover, ${L}(\hat{x})\rho$ = $\frac{1}{2}\left\{\hat{x}^\dag\hat{x},\rho \right\} - \hat{x}\rho\hat{x}^\dag$ is the Lindblad operator  %
and $[\hat{x},\hat{y}] =  \hat{x}\hat{y} -\hat{y}\hat{x}$, is the commutation relation, where $\hat{a}_k$ is creation operator for cavity mode $k$ and $\hat{\sigma}^\pm_i$ are the Pauli operators for the $i^{th}$ dye-molecule. 

The first term $\mathcal{L}^{(0)}[\rho]$ in the master equation captures the rate of loss of cavity photons $\kappa$, the de-excitation of molecules $\Gamma_\downarrow$ that creates photons outside the cavity modes, and the incoherent excitation of molecules $\Gamma_\uparrow$. The second term $\mathcal{L}^{(1)}$ arises from the incoherent coupling of different cavity modes with molecules via absorption and emission processes. This coupling term is given by, $\Psi_{k,k'}(\mathbf{r_i}) = \psi^*_{k}(\mathbf{r_i}) \psi_{k'}(\mathbf{r_i})$, where $\psi_{k}(\mathbf{r_i})$ is the wavefunction of cavity mode $k$, and $\mathcal{A}_k$ and  $\mathcal{E}_k$ are the absorption and emission rates.
It is important to note that although the above master equation accounts for the coupling between the optical modes, correlations between the molecules are neglected. This is due to the fact that the repeated collision of dye and solvent molecules lead to fast dephasing of any potential correlations between the molecules.

As discussed in the main text, the joint dynamics of the photon-molecule system is given by a set of rate equations derived from the above master equation.  At this point, the large number of dye-molecules in the system allows for the semiclassical approximation and correlations between photons and molecules can also be taken to be negligible, i.e., $\langle\hat{a}_k^\dag\sigma^-_i\rangle \approx \langle\hat{a}_k^\dag\rangle\langle\sigma^-_i\rangle$. The equation of motion for the photon correlation matrix $\textbf{n}$, with elements ${n}_{k,k'} = \langle \hat{a}_k^\dag\hat{a}_{k'}\rangle$, is given by Eq.~(\ref{eq:ndot}) in the main text.
%
In our numerical implementation, the  transverse 2D space of the cavity is divided into a finite number of bins (say $\mathcal{M}$), each located at $\mathbf{r_j}$.
%
The molecular excitation in the $j^{th}$ bin is given by $\mathbf{m}_j = \sum_{i} \delta(\mathbf{r_j}-\mathbf{r_i})\langle\sigma_i^+\sigma_i^-\rangle$. The equation of motion for the molecular excitation array $\mathbf{m}$ with $\mathcal{M}$ elements $m_j$, is then given by Eq.~(\ref{eq:fdot}).

The key parameters in our numerical simulation are the harmonic potential trap frequency $\wt = 0.5$~THz. The cavity cutoff is $\omega_0/\wt \approx 10^3$, the cavity photon loss is $\kappa/\wt \approx 0.26$, and the molecular de-excitation is $\Gamma_\downarrow/\wt = 0.002$. The incoherent pump rate is $\Gamma_\uparrow/\wt = 0.4$ and the density of molecules in the transverse plane is $3.12 \times 10^7$ (in units of $\ell_\mathrm{HO}^2$), where $\ell_\mathrm{HO}$ is the harmonic oscillator length discussed in Sec.~III.

\begin{figure*}[]
\includegraphics[width=0.99\textwidth]{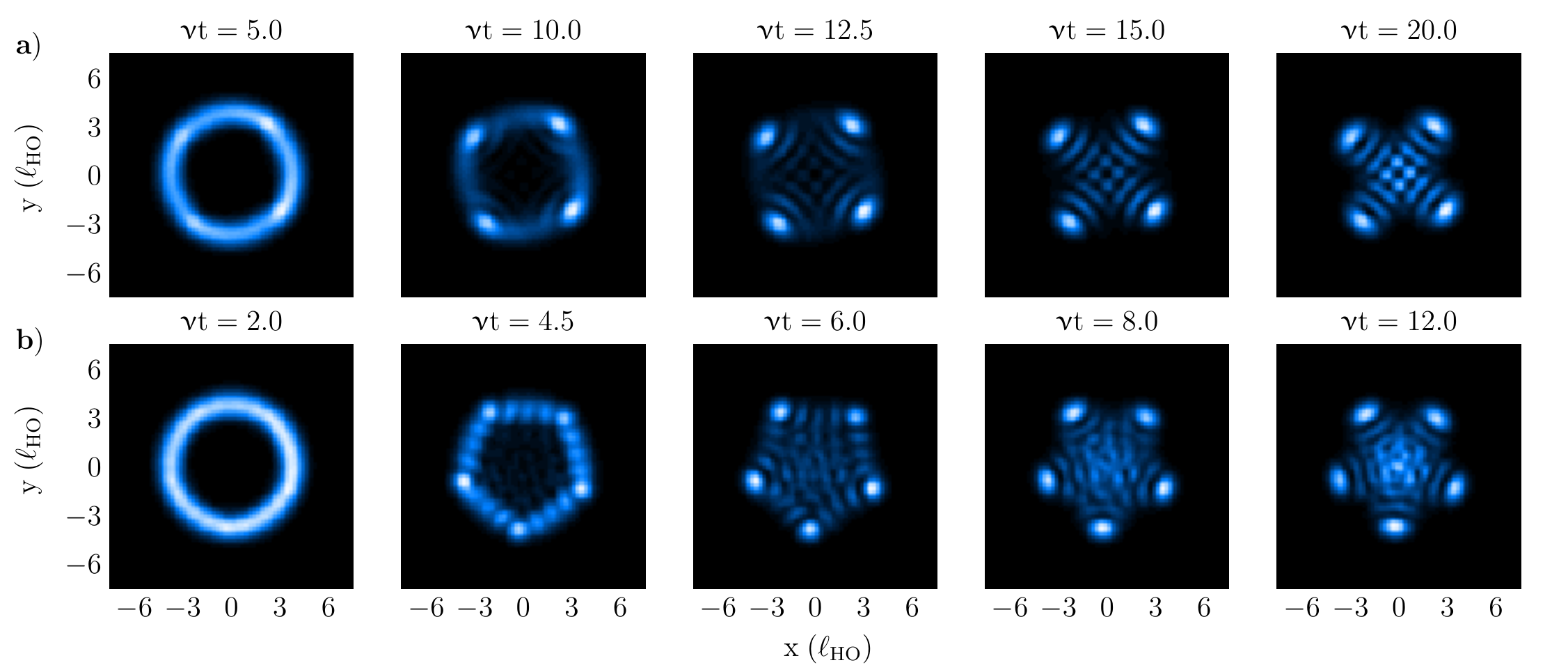}
\caption{Formation of transient spatial structures in condensed light. 
The figure shows both the initial and long time behavior of the photon condensate inside the cavity, in a frame co-rotating with the pump spot with orbital frequency $\nu$. After initial excitation, a circular ring of stimulated photon emission is observed. With increasing time, the spatial structure transforms to either a) a quadrilateral or b) a pentagon depending on the orbital pump frequency $\nu = \omega_\mathrm{T}/2$ or $ \omega_\mathrm{T}/5$, respectively, thus giving the condensate a discrete rotational symmetry. With further temporal evolution, the edges of the polygon start disappearing and the vertices morph into discrete spatial structures.  
}
\label{supp_fig0}
\end{figure*}
%

\begin{figure*}[]
\includegraphics[width=0.99\textwidth]{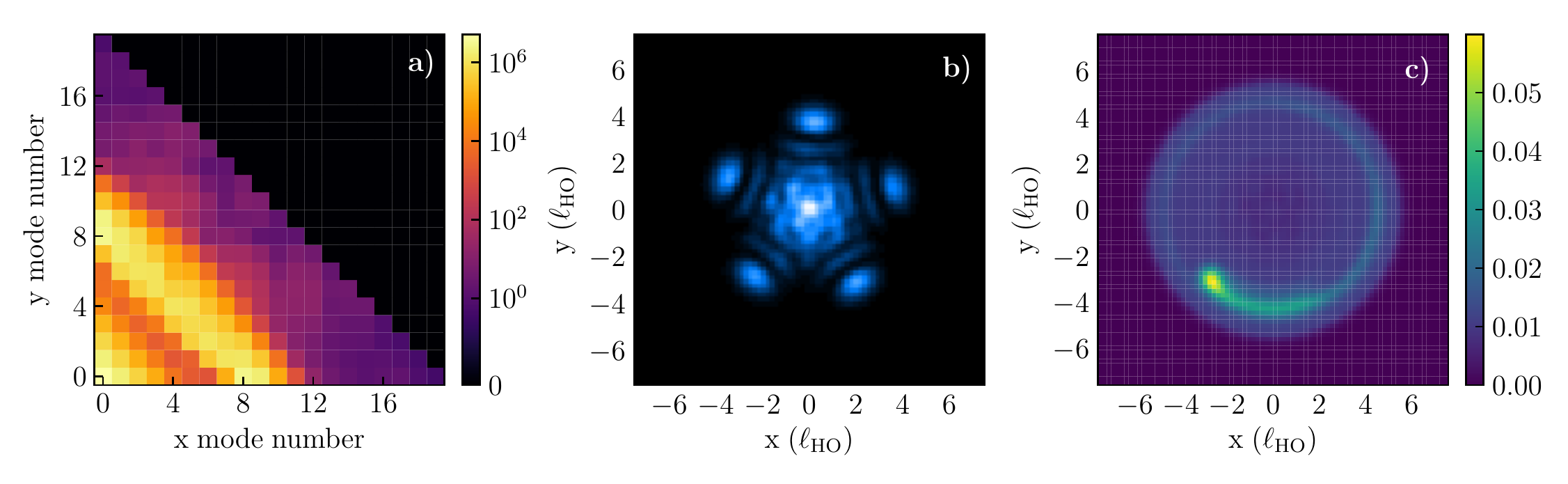}
\caption{A snapshot of the mode population, photon density and the molecular excitation inside a cavity driven by an incoherent pump with orbital frequency equal to $\wt/5$. The figure shows a) the occupation number of the different cavity modes ($x,y$) in the 2D transverse plane, b) the photon density of the condensed light and c) the fraction of excited molecules in each bin, at a fixed time. The photon population highlights the modes that participate strongly in forming the transient photon condensate, and therefore the resulting dynamical behavior as well as the discrete spatial structures.
}
\label{supp_fig2}
\end{figure*}

\subsection{II. Evolution of spatial structures of light}
\label{sec1}

In the main text we show the formation of phase-coherent, vortex-like spatial structures with discrete rotational symmetry. This occurs when the cavity is driven by an incoherent, focused pump, with the pump spot orbiting the cavity plane with a frequency $\nu$, which is a fraction of the frequency of the harmonic trap $\omega_\mathrm{T}$. In other words, the photon densities with discrete structures are seen when $\nu = \omega_\mathrm{T}/z$, where $z$ is an integer. 

Importantly, the discrete rotational symmetry gradually appears as the system evolves with time, with the initial spatial structures undergoing several key transformations.
We highlight this in Fig.~\ref{supp_fig0} of the Supplemental Material. At initial times, when the molecules are excited in the annular region along the pump orbit, the photon emission gives rise to a ring-like structure that gradually transforms to a regular polygon.  
In a frame co-rotating with the pump spot, the transient spatial structures appear stationary and slowly adopt a distinct odd-even dichotomy depending on the value of $z$. For even and odd $z$, a polygon with $2z$ and $z$ vertices are observed, respectively. For instance, in Fig.~\ref{supp_fig0}, for $z=2$ a quadrilateral is observed, whereas a pentagon  is observed for $z=5$. As $t$ increases, the edges of the polygon disappear and the vertices morph into discrete peaks, with interference patterns between them, and a distinct four-fold or five-fold rotational symmetry 
is visible in the condensed light.

\subsection{III. Interference of cavity modes}

For stationary pumping or a fixed pump spot, an initial photon wavepacket is formed due to stimulated emission, which is a nonequilibrium condensate formed by the superposition of several cavity modes. Such a photon wavepacket oscillates inside the cavity with the frequency $\wt$ of the harmonic potential that traps the photon gas.
An important question discussed in the main text is about the dominant modes that contribute to this oscillating condensed light. 
This typically depends on the distance of the pump spot from the cavity center or the radius of orbit of the circulating pump spot when in motion. Now, for a one-dimensional quantum harmonic oscillator, the most probable region occupied by a particle with energy $E_n$ or the $n^{th}$ mode is given by $A_n \approx \sqrt{2n\frac{h}{m_\textrm{ph}\wt}}$, where $m_\textrm{ph} = h\omega_0/c^2$ is the effective photon mass. Here, $\omega_0$ is the cavity cutoff frequency and $c$ is the speed of light in vacuum. The quantity $\sqrt{\frac{h}{m_\textrm{ph}\wt}}$ is the harmonic oscillator length $\ell_{\mathrm{HO}}$.  If the focused pump now excites molecules at a distance $r$ from the cavity center, the relevant excited modes are those that correspond to $n$ = $r^2/(2 \ell_{\mathrm{HO}}^2)$. 
For 2D harmonic oscillators, the frequency of mode $k$ is given by $\omega_k = \omega_0 + (q_x+q_y)\wt$, corresponding to the set of 1D quantum numbers $\{q_x,q_y\}$.
Therefore, the most probable modes here are those that are characterized by the condition $q_x + q_y$ =  ${r}^2/(2 \ell_\textrm{HO}^2)$.

Figure~\ref{supp_fig2}, shows the mode occupation, photon emission and the molecular excitation fraction at a fixed time, for an incoherent pump with orbital frequency $\wt/5$. The photon density in Fig.~\ref{supp_fig2}(b) shows a five-fold rotational symmetry with discrete peaks. The interference pattern is distinct and there is  occupation of the ground state at the center of the spatial structure. The incoherent pump spot is $r = 4 \ell_\mathrm{HO}$, which is evident from the annular region of molecular excitation fraction in Fig.~\ref{supp_fig2}(c). Now the relevant modes that form the photon wavepacket is $q = {r}^2/(2 \ell_\textrm{HO}^2) = 8$. This is supported by the mode occupation in Fig.~\ref{supp_fig2}(a), where the modes close to those with energy $q = 8$ are highly occupied, along with the lowest lying energy states. While the former forms the discrete peaks and the rotationally symmetric structure, the latter forms the bright photon intensity at the center of the spatial structure. The condensed light here splits into a nonequilibrium condensate co-rotating with the pump spot in the cavity plane and a near-equilibirum Bose-Einstein condensate at the center.


\begin{thebibliography}{99}
\bibitem{comment1} For $z$ that are not integers, there is a qualitative trade-off between the amount by which $z$ is non-commensurate and the time over which spatial structures are robust. For $z$ close to an integer, the spatial structures remain distinctly visible over a substantially longer period, in contrast to the opposite case where the discrete structure becomes unclear relatively early.
\end{thebibliography}
\end{document}